\documentclass[a4paper,fleqn]{cas-sc}

\usepackage[authoryear,longnamesfirst]{natbib}
\usepackage{siunitx}
\usepackage{booktabs}
\usepackage{tabularx}
\usepackage{threeparttable}
\usepackage{longtable}
\usepackage{array}
\usepackage{pdflscape}
\usepackage{graphicx}
\usepackage{wrapfig}
\usepackage{needspace}
\usepackage{amsmath}
\usepackage{amssymb}
\def\tsc#1{\csdef{#1}{\textsc{\lowercase{#1}}\xspace}}
\tsc{WGM}
\tsc{QE}

\begin{document}
\let\WriteBookmarks\relax
\def\floatpagepagefraction{1}
\def\textpagefraction{.001}

\shorttitle{}    

\shortauthors{}  

\title [mode = title]{Feasibility of the Phobos 1 Hypothesis for Dark Comet 1998 KY$_{26}$}  

%

\author[1]{Adam Hibberd}[orcid=0000-0003-1116-576X]
\cormark[1]
\cortext[1]{Corresponding author.}
\ead{adam.hibberd@i4is.org}

\author[1]{Adam Crowl}[orcid=0009-0009-3118-8513]
\author[2]{Abraham Loeb}[orcid=0000-0003-4330-287X]

\credit{Adam Hibberd}

\affiliation[1]{organization={Initiative for Interstellar Studies},
            city={27/29 South Lambeth Road London},
            postcode={SW8 1SZ}, 
            state={London},
            country={United Kingdom}}
\affiliation[2]{organization={Astronomy Department,Harvard University}, addressline={60 Garden Street}, city={Cambridge}, postcode = {MA 02138}, country={USA}}

\begin{abstract}
The asteroid 1998 KY$_{26}$ has been the subject of thorough observation, both optical and radar, from shortly following its discovery on 28 May 1998, through 2 further close apparitions in 2020 and 2024. This has previously allowed an accurate characterization of the object, including an unusually rapid spin-rate with period 5.3516 $\pm$ 0.0001 minutes, a diameter of 11 $\pm$ 2 m, and a high albedo of $\sim{0.52}$. Furthermore, the presence of significant nongravitational accelerations (NGAs), with no detectable shedding of gas or dust, has stimulated the 'dark comet' categorization, with JAXA repurposing their Hayabusa2$\sharp$ spacecraft to rendezvous with the object in 2031. We follow-up the astrodynamical evidence pointing to the possibility this may actually be the lost Soviet Phobos 1 probe, and analyse various photometry and astrometry associated with 1998 KY$_{26}$ to further investigate the feasibility of this hypothesis. We find a first order approximation of the Phobos 1 spacecraft provides compelling agreement to 8 light curves of the object and further that modelling NGAs as solar radiation pressure (SRP) on a cylinder or solar panels provide significant reductions in residual of $\sim{11} \%$ with respect to the 260 astrometric and radar measurements. Although an investigation of this kind cannot be conclusive, nevertheless the results here add strong weight to the Phobos 1 hypothesis, and we find no clear contradictory evidence. The spin pole in Ecliptic J2000 coordinates, calculated from photometry based on the Phobos 1 assumption is $(\lambda,\beta)=({151^{+1}_{-2}}^{\circ},{+11^{+12}_{- 3}}^{\circ})$ , where the quoted ranges represent approximate 68.3\% confidence levels.    
\end{abstract}


\begin{highlights}
\item Photometry, radar and astrometry of the dark comet 1998 KY$_{26}$ are investigated in the context of the Phobos 1 probe hypothesis, finding supporting evidence.
\item There is general agreement between first order approximations of the Phobos 1 probe and observations.
\item Small differences between the expected light curves and actual light curves are present and manifest as odd harmonics, maybe due to the problem of fitting a complicated spacecraft shape with as few parameters as possible.   
\end{highlights}

\begin{keywords}
1998 KY$_{26}$ \sep Phobos 1 \sep Dark Comets\sep
\end{keywords}

\maketitle

\section{Introduction}\label{INTRO}

Discovered on 28 May 1998 UT \citep{MPC1998KY26,Spacewatch1998KY26}, the asteroid 1998 KY$_{26}$ was quickly tracked by optical observations (2-8 June) and radar (6-8 June). Immediately following this \cite{Hicks1998CloseEncounters} embarked on an early physical characterization of 1998 KY$_{26}$, calculating a rotation period of  10.69 $\pm$ 0.02 $\si{mins}$, leading to a size estimate from radar measurements of $\sim{40}$ $\si{m}$. According to \cite{Hicks1998CloseEncounters}, the high rotation rate indicated a  monolithic body, since a rubble-pile should fly apart at such high rates. Table \ref{tab:ky26_observation_stretches} provides all the optical observations to date of 1998 KY$_{26}$. The optical observations used by \cite{Hicks1998CloseEncounters} are referenced '1998 discovery' in this Table.\\

A more thorough analysis of these discovery observations was conducted by \cite{Ostro1999KY26}. Analysis of the light curve by this team revealed an unambiguous time period of $P =$ 10.7015 $\pm$ 0.0004 $\si{mins}$. From this period $P$, and exploiting radar measurements, an approximate diameter could be calculated using measured full Doppler bandwidth, $B$, and an estimate of the latitude of the line-of-sight with respect to the spin plane of the object, $\delta$. This led \cite{Ostro1999KY26} to derive an estimate for 1998 KY$_{26}$ of $\sim{30}$ $\si{m}$.\\

It should be noted that this time period was later to be questioned by subsequent analysis with additional astrometry, eventually favouring a more probable period of half that of the original research (this time period is reproduced in the research herein), thus scaling the size down by a factor of around two. Nonetheless, as a result of the 10.7 minute premise by \cite{Ostro1999KY26}, and with its high absolute magnitude of H $\sim{25.5}$, they concluded a low visual albedo satisfying $0.05 \le p_v \le 0.37$, naturally implying a relatively dark carbonaceous chondrite asteroid. \\

\begin{wrapfigure}{l}{0.62\textwidth}
\centering
\includegraphics[width=0.6\textwidth]{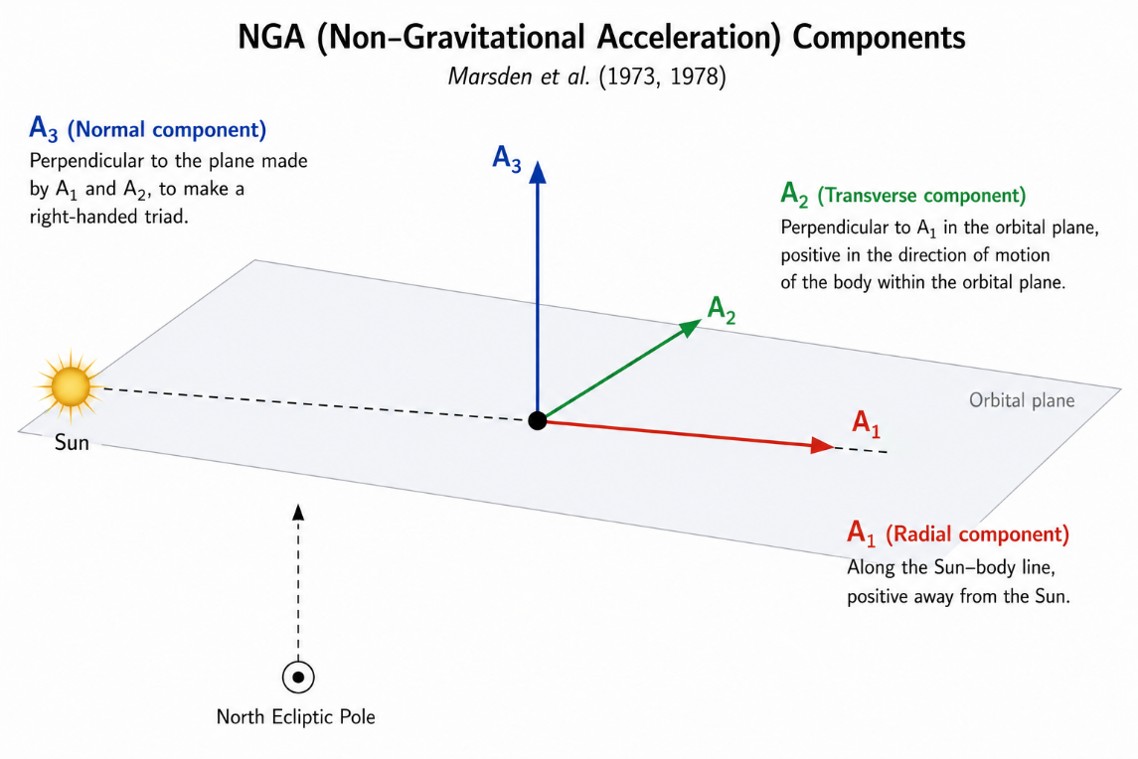}
\caption{Depiction of the three nongravitational accelerations (NGAs), (A$_1$, A$_2$, A$_3$), the diagram is self-explanatory.}
\label{fig:Marsden}
\end{wrapfigure}

The 'dark comet' hypothesis for 1998 KY$_{26}$ was posited in \cite{Seligman2023DarkComets}, as a consequence of the additional observations labelled '2020 recovery' in Table \ref{tab:ky26_observation_stretches}. The paper provides an estimate of the 3 so-called 'nongravitational acceleration' (NGA) components of 1998 KY$_{26}$, assuming the model used by \cite{Marsden1973}. Refer Figure \ref{fig:Marsden} for the definition of the three NGA components (A$_1$, A$_2$, A$_3$)\\

In theory, under the gravitational influence of the Sun only and with no perturbing forces, the 5 'Keplerian orbital elements' should remain fixed over time. Of course this ideal is not exactly respected in our solar system where the gravitational forces of the planets, especially the Gas Giants Jupiter and Saturn for example, create perturbations in the motion of a body, leading to gradual changes in its 'osculating orbital elements'.\\

\Needspace{0.65\textheight}

Since we know the Ephemerides of all the planets as a function of time extremely accurately, we can quite easily and precisely compensate for these perturbations and still construct an accurate prediction of any body's future path under ALL these gravitational forces. However, it is when this fully gravity-accounted for path diverges from prediction that we must evoke NGAs.\\

Possible causes include:
\begin{enumerate}
    \item Outgassing of volatile compounds from a comet's surface as they heat up approaching the Sun
    \item Forces due to solar radiation and rotation of the body axes, the Yarkovsky Effect
    \item Solar radiation pressure (SRP)
\end{enumerate}

In the following report we shall examine item (3) above for 1998 KY$_{26}$. Item (1) implies the possibility of jets of gas or dust ejected from the object causing accelerations from the 'rocket effect'. For the generalized model proposed by \cite{Marsden1973}, we have:

\begin{equation}
    \mathbf{a_{NG}} = g(r)( A_1 \hat{\mathbf{r}} + A_2 \hat{\mathbf{t}} + A_3 \hat{\mathbf{n}} )
\end{equation}

Where $\hat{\mathbf{r}}$, $\hat{\mathbf{t}}$ and $\hat{\mathbf{n}}$ are unit vectors in the Sun-radial, transverse and normal directions - respectively the red, green and blue arrows in Figure \ref{fig:Marsden}. For a comet, i.e. in the case of item (1) listed above, the form of the Sun-radial normalization $g(r)$ depends on the volatiles; but for items (2) and (3), which are the main contribution to NGAs for an asteroid, then we can take $g(r) = au^2/r^2$, reflecting the inverse square dependence of these forces on solar distance.\\

As a consequence of the work conducted by \cite{Seligman2023DarkComets,Seligman2024DarkCometPopulations}, 9 objects shown to have no cometary behaviour, with no evidence of outgassing, were nevertheless found to have significantly non-zero NGAs, particularly in the transverse (A$_2$) and normal directions (A$_3$). Since A$_3$ is not at all unusual in the outgassing scenario, but strange for solar radiation pressure on an irregularly shaped object such as an asteroid, the conclusion of \cite{Seligman2023DarkComets} was that the former activity must be taking place on these objects, despite there being no photometric evidence, hence the proposed new category of object: the 'dark comet'. \\

With further optical observations of 'dark comet' 1998 KY$_{26}$ during its apparition in 2024 (refer Table \ref{tab:ky26_observation_stretches} - 2024 apparition), 3 papers ensued \cite{Beniyama2025KY26Size,Bolin2025KY26,SantanaRos2025KY26}. In particular the latter determined that the light curve of the object was highly regular, and the spin rate precisely half that determined previously by \cite{Ostro1999KY26}, at 5.3516 $\pm$ 0.0001 $\si{mins}$. This impacted on the interpretation of the radar sightings in 1998, implying a much smaller diameter of 11 $\pm$ 2 $\si{m}$. This also had an appreciable effect on the visual albedo, indicating $p_v$ $\sim{0.52}$, suggesting a shiny and hard 'Xe'-type monolithic object as opposed to the previous carbonaceous chondrite assumption.\\

The light curve analysis of \cite{SantanaRos2025KY26} allowed them to compute a rotation pole of the object in ecliptic coordinates, and furthermore this pole was exploited by \cite{Farnocchia2025KY26} to ascertain a best fit oblate spheroid shape to the object, on the basis that the forces causing the significant NGAs on 1998 KY$_{26}$ were due to solar radiation - principally SRP, but also Yarkovsky. Their model resulted in an excellent orbital fit, superior to the Marsden model, and demonstrated that no additional outgassing was necessary.\\
\Needspace{0.4\textheight}
\begin{wrapfigure}{l}{0.46\textwidth}
\centering
\includegraphics[width=0.45\textwidth]{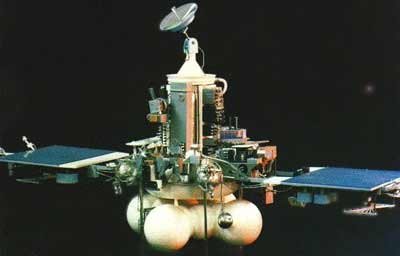}
\caption{Photograph of Phobos 1 probe in cruise configuration with ADU attached - the 8 white globes at the bottom.}
\label{fig:PhobosPhoto}
\end{wrapfigure}
This finding is relevant since the JAXA (Japanese Aerospace Exploration Agency) is repurposing its 'Hayabusa2$\sharp$' spacecraft mission to rendezvous with 1998 KY$_{26}$ in July 2031. Its observations shall finally settle the matter of the ontology of this curious object, which has generated considerable debate, especially in the light of a new theory suggesting the object might be the Soviet Phobos 1 probe - refer Figure \ref{fig:PhobosPhoto} - lost on its outward journey to Mars back in September of 1988 \citep{Hibberd2026Phobos1}.\\

The present paper looks into this latter hypothesis in more detail and is intended as a feasibility study, examining the consequences of this hypothesis on the expected behaviour of the object and revealing compelling evidence to support such an identity for 1998 KY$_{26}$. It should be noted however, that it canNOT unambiguously prove this identity.

\section{Light Curve Analysis}\label{sec:lc}

In this section we adopt the optimal time period of 5.3516 $\si{mins}$ fitted by \cite{SantanaRos2025KY26} and conduct further Fourier Analysis on the same reduced photometric source data used by \cite{SantanaRos2025KY26} paper. They found no secondary periodicity in the light curve (which would otherwise be indicative of nutation) and found a principal axis rotation with a fixed pole of (36$^{\circ}$, -44$^{\circ}$) in ecliptic coordinates. The data comprises 954 photometric measurements and 22 source segments, in turn organized into 8 observation groups, each group with solar phase angles varying between 5$^{\circ}$ and 101$^{\circ}$.\\

\begin{figure}
\centering
\includegraphics[width=1.0\textwidth]{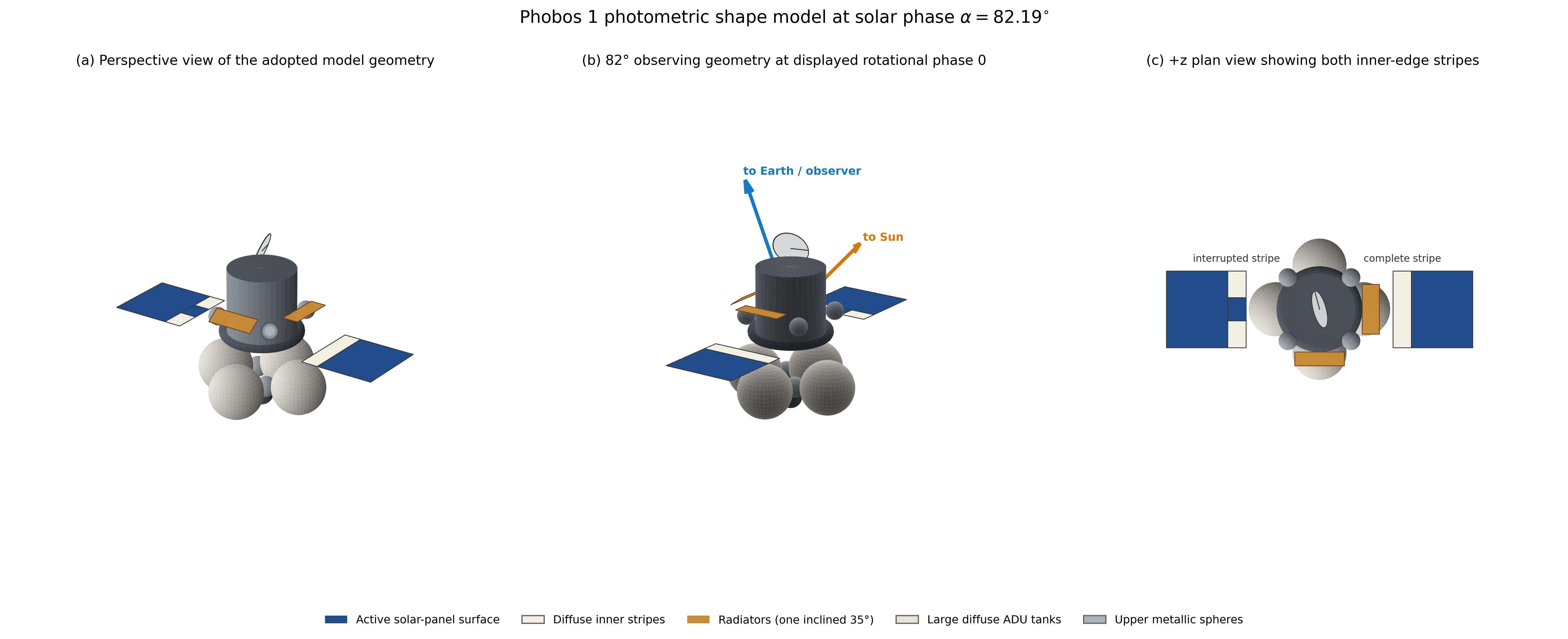}
\caption{The eventual shape and structure of the modelled spacecraft as a result of an iterative process to match 8 light curves for 1998 KY$_{26}$. The precise dimensions are largely optimal best-fit parameters and are close to those of the Phobos 1 probe adding significant weight to the hypothesis.}
\label{fig:Approx}
\end{figure}

The research covered in this section cannot uniquely identify 1998 KY$_{26}$ as the Phobos 1 probe, and is also not an attempt to fit absolute magnitudes to the dark comet, indeed each group is normalized so that only changes in magnitude within a group are fitted. A shape model for the Phobos 1 probe is constructed (refer Appendix \ref{app:pole_search} for the detail). The amplitude and shape of the waveform is modelled and compared. The rotational phase offset (or the 'roll angle') is refitted for each observation group, thus rotational phase synchrony across a wide range of epochs is not assumed. A photograph of the Phobos 1 probe is provided in Figure \ref{fig:PhobosPhoto}, it is highly irregular and asymmetrical. After several iterations, the Phobos 1 probe approximation, including its ADU Mars insertion module, converged to that shown in Figure \ref{fig:Approx}. 

\begin{landscape}
\begin{table}[p]
\centering
\caption{Observation stretches for 1998 KY26 derived from the MPC-format
astrometric file.}
\label{tab:ky26_observation_stretches}

\resizebox{\linewidth}{!}{%
\begin{tabular}{llllllr}
\toprule
Apparition & Start date & End date & MPC & Observatory or station & Type & Records \\
\midrule
1998 discovery & 1998-05-28 & 1998-06-05 & 691 & Steward Observatory, Kitt Peak--Spacewatch & Optical & 18 \\
1998 discovery & 1998-05-31 & 1998-06-06 & 046 & Kleť Observatory, České Budějovice & Optical & 48 \\
1998 discovery & 1998-05-31 & 1998-06-07 & 557 & Ondřejov Observatory & Optical & 44 \\
1998 discovery & 1998-06-01 & 1998-06-01 & 860 & Valinhos Observatory & Optical & 3 \\
1998 discovery & 1998-06-01 & 1998-06-03 & 709 & W \& B Observatory, Cloudcroft & Optical & 13 \\
1998 discovery & 1998-06-01 & 1998-06-06 & 422 & Loomberah & Optical & 15 \\
1998 discovery & 1998-06-01 & 1998-06-07 & 118 & Astronomical and Geophysical Observatory, Modra & Optical & 11 \\
1998 discovery & 1998-06-02 & 1998-06-02 & 817 & Sudbury Observatory & Optical & 4 \\
1998 discovery & 1998-06-02 & 1998-06-02 & 784 & Stull Observatory, Alfred University & Optical & 6 \\
1998 discovery & 1998-06-02 & 1998-06-06 & 426 & Woomera & Optical & 10 \\
1998 discovery & 1998-06-03 & 1998-06-03 & 704 & Lincoln Laboratory ETS, New Mexico & Optical & 5 \\
1998 discovery & 1998-06-04 & 1998-06-04 & 327 & Peking Observatory, Xinglong Station & Optical & 3 \\
1998 discovery & 1998-06-05 & 1998-06-05 & 595 & Farra d'Isonzo & Optical & 3 \\
1998 discovery & 1998-06-05 & 1998-06-05 & 670 & Camarillo & Optical & 7 \\
1998 discovery & 1998-06-05 & 1998-06-05 & 568 & Maunakea & Optical & 5 \\
1998 discovery & 1998-06-06 & 1998-06-06 & 686 & University of Minnesota Infrared Observatory, Mt Lemmon & Optical & 10 \\
1998 discovery & 1998-06-06 & 1998-06-08 & 658 & Dominion Astrophysical Observatory & Optical & 9 \\
1998 discovery & 1998-06-06 & 1998-06-08 & 252/253 & Goldstone DSS-13/DSS-14 & Radar & 6\textsuperscript{a} \\
\midrule
2002 recovery & 2002-02-16 & 2002-02-17 & T12 & University of Hawaii 88-inch Telescope, Maunakea & Optical & 2 \\
\midrule
2020 recovery & 2020-12-10 & 2020-12-12 & 309 & Cerro Paranal/VLT & Optical & 5 \\
2020 recovery & 2020-12-10 & 2020-12-10 & 568 & Maunakea/Subaru & Optical & 3 \\
2020 recovery & 2020-12-15 & 2020-12-15 & 950 & La Palma & Optical & 10 \\
\midrule
2024 apparition & 2024-04-17 & 2024-04-17 & 309 & Cerro Paranal/VLT & Optical & 5 \\
2024 apparition & 2024-05-01 & 2024-05-01 & G37 & Lowell Discovery Telescope & Optical & 2 \\
2024 apparition & 2024-05-06 & 2024-05-06 & F51 & Pan-STARRS 1, Haleakala & Optical & 3 \\
2024 apparition & 2024-05-11 & 2024-05-11 & J13 & Liverpool Telescope, La Palma & Optical & 1 \\
2024 apparition & 2024-05-19 & 2024-05-19 & W84 & Cerro Tololo--DECam & Optical & 4 \\
2024 apparition & 2024-05-21 & 2024-05-21 & Z18 & Gran Telescopio Canarias & Optical & 4 \\
2024 apparition & 2024-05-29 & 2024-05-29 & 691 & Spacewatch, Kitt Peak & Optical & 2 \\
2024 apparition & 2024-06-03 & 2024-06-03 & G37 & Lowell Discovery Telescope & Optical & 2 \\
\bottomrule
\end{tabular}%
}
\vspace{2mm}
\begin{minipage}{0.96\linewidth}
\footnotesize
\textsuperscript{a}The six radar records correspond to three bistatic
Goldstone epochs on 1998 June 6, 7 and 8. Each epoch contains paired
range and Doppler records. DSS-14 transmitted and DSS-13 received.
\end{minipage}

\end{table}
\end{landscape}




\begin{figure}
\centering
\includegraphics[width=1.0\textwidth]{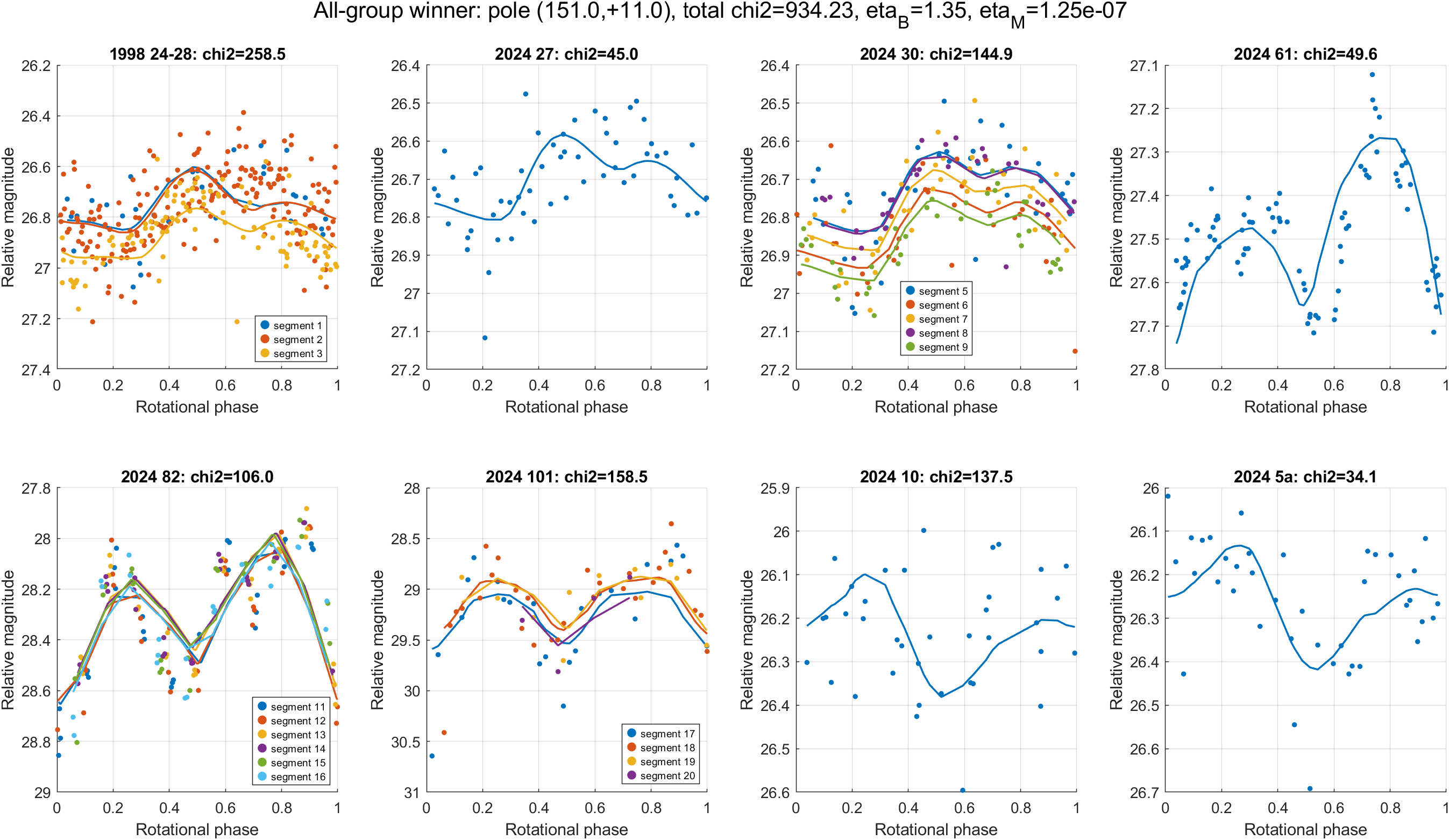}
\caption{8 of the 9 groups of light curves (with the groups delineated by common phase angle, $\alpha$) used by \cite{SantanaRos2025KY26} to establish the 5.3516 $\si{min}$ rotation period, with the model of Phobos 1 as solid lines. Note the 2024 June 5 observations are labelled as having a phase angle of $(87^\circ)$ by \cite{SantanaRos2025KY26}; however, calculation from the Sun and Earth vectors supplied with their photometric source data gives $(\alpha\simeq82.2^\circ)$, consistent with JPL Horizons.}
\label{fig:LightCurves}
\end{figure}

The principal optical and geometrical assumptions of the model were:
\begin{enumerate}
    \item Lambertian diffuse reflection from the solar panels, central bus, 
          white inner-edge panel stripes, large lower ADU tanks, and the 
          reflective face of each radiator;
    \item opaque surfaces, with mutual shadowing and observer-side occultation;
    \item infinitesimally thin rectangular solar panels, with a finite active 
          breadth and inner radius;
    \item a complete bright stripe along the inner edge of one solar panel and 
          a corresponding stripe with a central interruption on the opposite 
          panel;
    \item two rectangular radiators separated by $90^\circ$ in azimuth, with 
          their long axes tangential to the central bus and therefore 
          perpendicular to their respective radial position vectors;
    \item one radiator was inclined to the solar-panel plane, with both faces assigned the same effective diffuse-reflection coefficient in the implemented model;
    \item a flat circular approximation to the parabolic high-gain antenna, 
          with effective brightness equivalent to q$_{bus}$ on its rear face;
    \item spherical or sampled-sphere approximations for the large lower ADU 
          tanks, smaller attitude-control tanks, and components approximating 
          the toroidal spacecraft structure;
    \item an optional parametrised specular contribution from the upper 
          metallic attitude-control spheres, represented using sampled 
          spherical surfaces and a fixed specular exponent;
    \item no thermal-emission contribution; and
    \item no complete spacecraft CAD model or representation of small-scale 
          structural components.
\end{enumerate}

An initial raster scan was performed over the sphere of possible spin poles, 
specified by the ecliptic J2000 longitude and latitude $(\lambda,\beta)$. 
Successive local pole scans were then used to refine the preferred region. 
For each trial pole, the photometric model included the following fitted or 
profiled parameters:
\begin{enumerate}
    \item $\gamma_{\rm ant}$, the clock angle of the high-gain antenna, during 
          the initial geometry searches;
    \item $q_{\rm bus}$, the effective brightness of the bus relative to the 
          active solar-panel surface;
    \item $q_{\rm ant}$, the effective brightness of the antenna relative to 
          the active solar-panel surface;
    \item $\eta_{\rm board}$, the radiator brightness relative to the bus, such 
          that
          \[
              q_{\rm board}=\eta_{\rm board}q_{\rm bus};
          \]
    \item $\eta_{\rm stripe}$, the brightness of the white inner-edge stripes 
          relative to the adopted reference surface;
    \item $\eta_{\rm metal}$, the relative strength of the optional 
          metallic-sphere reflection component;
    \item $\phi_g$, an independent rotational-phase offset for each of the 
          eight fitted observation groups; and
    \item $Z_j$, $j=1,\ldots,22$, an additive magnitude zero point for each 
          observational segment.
\end{enumerate}

The solar-panel breadth and inner radius, together with the inclination of the 
second radiator, were investigated in separate geometrical scans before the 
final pole refinement. In particular, reproducing the observed widths and 
locations of important light-curve features required an active panel breadth 
close to $2.5\,\mathrm{m}$, consistent with the value estimated independently 
from photographs of Phobos~1. The radiator-inclination scan favoured an angle 
of approximately $35^\circ$.\\

The final local pole scan held the antenna clock angle at 
$\gamma_{\rm ant}=195^\circ$, the radiator inclination at $35^\circ$, and the 
active panel breadth at $2.5\,\mathrm{m}$. The quantities $q_{\rm bus}$, 
$q_{\rm ant}$, $\eta_{\rm board}$ and $\eta_{\rm metal}$ were profiled against 
the combined $\chi^2$ of all eight observation groups. The preferred solution 
was located at:
\begin{equation}
      (\lambda,\beta)=({151^{+1}_{-2}}^{\circ},{+11^{+12}_{- 3}}^{\circ}),  
\end{equation}

where the quoted ranges are the approximate 68.3\% projected confidence extremes.
The quantities $q_{\rm bus}$, $q_{\rm ant}$, $q_{\rm board}$, 
$\eta_{\rm stripe}$ and $\eta_{\rm metal}$ are effective photometric 
coefficients rather than measured material albedos. In the final all-group 
solution, $\eta_{\rm metal}$ converged to a value effectively indistinguishable 
from zero, indicating that an explicit specular contribution from the upper 
metallic spheres was not required by the fitted light curves.\\

\Needspace{0.50\textheight}
\begin{wrapfigure}{l}{0.50\textwidth}
\centering
\includegraphics[width=0.5\textwidth]{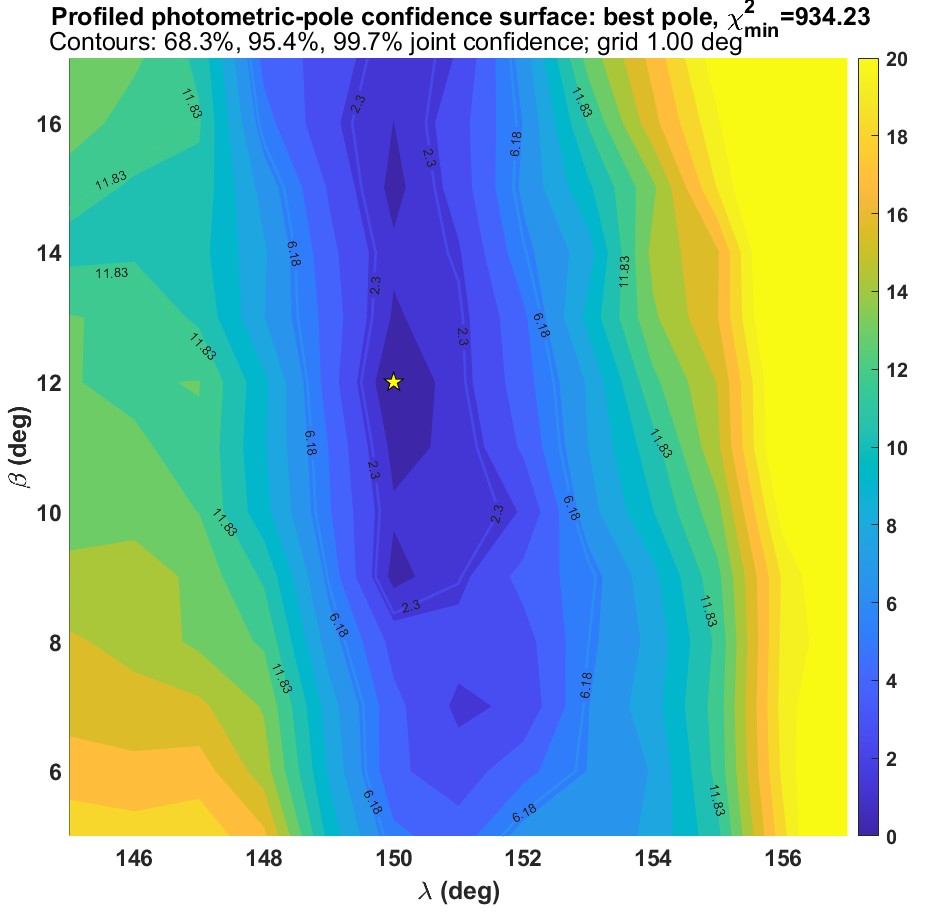}
\caption{Contours of $\Delta \chi^2$ relative to the minimum of $\chi^2=934.23$, from which the uncertainties in $(\lambda, \beta)$ can be derived}
\label{fig:ContoursPoles}
\end{wrapfigure}
After several iterations of the model, the results provided here are from the most sophisticated version described in Appendix \ref{app:pole_search}. \\

Figure \ref{fig:LightCurves} show general agreement, however there are some features absent from the fit at high solar phase angles of $\sim{82^{\circ}}$ and $\sim{101^{\circ}}$. This could be due to aspects of Phobos 1 remaining unmodelled, such as the irregularity of the central bus, yielding an over-simplified representation of the spacecraft.\\

Refer to Figure \ref{fig:ContoursPoles} for a contour plot showing the optimal pole (the star) with $\chi^2=934.23$, and contours of $\Delta \chi^2$ around this optimum. This plot was used to determine the approximate 68.3\% uncertainty in the pole coordinates $\lambda$ and $\beta$.\\

For the preferred pole $(\lambda,\beta)=(151^\circ,+11^\circ)$, the model gives $\chi^2=934.23$ for 954 photometric measurements. Taking the 22 segment magnitude offsets, eight observing-group phase offsets and four global photometric coefficients as fitted parameters gives approximately $\nu=954-34=920$ degrees of freedom and hence a reduced $\chi^2$ of $\chi_\nu^2=934.23/920\simeq1.02$. Counting the two pole coordinates as additional fitted quantities changes this only slightly, to $\chi_\nu^2\simeq1.018$. A reduced $\chi_\nu^2$ so close to unity indicates that the typical model residual is approximately equal to the empirically estimated observational uncertainty; equivalently, the root-mean-square normalized residual is only $\sqrt{\chi_\nu^2}\simeq1.01$ standard deviations. The model therefore provides a statistically close description of the complete photometric dataset under the adopted error model, with no evidence from the overall $\chi^2$ for a substantial global mismatch. This does not establish that the spacecraft interpretation is unique, nor does it exclude coherent residual features within individual light curves, but it demonstrates that the approximated Phobos~1 geometry reproduces the observations to approximately the level permitted by their estimated scatter.\\

Two controls were performed:
\begin{enumerate}
    \item Just two solar panels rotating about a pole.
    \item Two solar panels rotating about a pole with a central cylindrical bus.
\end{enumerate}

Number (1) above should give a flat line to each of the 8 light curves in turn, and is a useful general reference for the the full Phobos 1 model. Control (2) allows an assessment of how shadowing of the panels by the bus and occultation by and of the bus improve the $\chi^2$ fit, compared to (1).\\

The results of these control experiments are shown in Table \ref{tab:photometric_controls}. They reveal that there is a 48\% improvement in $\chi^2$ fit by introducing the bus to the solar panels only geometry, and a further 52\% improvement by incorporating the various additional paraphernalia as mentioned previously in this section. Furthermore, when the pole is recalculated for the case (2) above, the optimal pole is within only 1.4$^{\circ}$ of the full Phobos 1 model, demonstrating that this pole is governed largely by this bus/panel geometry.\\

\begin{table}[t]
\centering
\caption{Comparison of the photometric control models with the full
Phobos~1 approximation.  The incremental $\Delta\chi^2$ is measured
relative to the preceding model.  The final column gives the fraction of
the total improvement from the flat two-panel null model to the full
Phobos~1 model.}
\label{tab:photometric_controls}
\small
\setlength{\tabcolsep}{5pt}
\renewcommand{\arraystretch}{1.20}
\begin{tabular}{lcccc}
\toprule
\textbf{Model}
& \textbf{Best pole $(\lambda,\beta)$}
& \textbf{Total $\chi^2$}
& \textbf{Incremental $\Delta\chi^2$}
& \textbf{Improvement fraction} \\
\midrule
Two Lambertian panels only
& N/A
& $1506.42$
& ---
& --- \\

Two panels $+$ cylindrical bus
& $(150^\circ,+12^\circ)$
& $1231.65$
& $274.77$
& $48.0\%$ \\

Full Phobos~1 approximation
& $(151^\circ,+11^\circ)$
& $934.23$
& $297.42$
& $52.0\%$ \\
\midrule
Total improvement over flat-panel null
& ---
& ---
& $572.19$
& $100.0\%$ \\
\bottomrule
\end{tabular}

\vspace{2pt}
\begin{minipage}{0.96\linewidth}
\footnotesize
The panels-only model is rotationally invariant and therefore has no
defined photometric pole.  The panels-plus-bus control includes diffuse
reflection from the cylindrical bus and bus shadowing and occultation of
the panels.  The full approximation additionally includes the antenna,
panel stripes, radiators, tanks, toroidal structure and the other
Phobos~1-specific components.  All three models use the same
photometric observations, statistical weights and independently fitted
segment magnitude zero-points.  The angular separation between the
control-model and full-model poles is approximately $1.4^\circ$.
\end{minipage}
\end{table}

\section{Estimate of Solar Radiation Pressure on Solar Panels}\label{}

\begin{table}
\centering
\caption{Marsden nongravitational-acceleration coefficients for
1998 KY$_{26}$ reported by \cite{Seligman2024DarkCometPopulations}. The coefficients are
normalized at a heliocentric distance of 1 au.}
\label{tab:marsden_coefficients}
\begin{tabular}{clccc}
\toprule
Coefficient &
Direction &
Value (\(\mathrm{au\,d^{-2}}\)) &
Value (\(\mathrm{m\,s^{-2}}\)) &
Significance \\
\midrule
\(A_1\) &
Radial &
\((1.60 \pm 0.88)\times10^{-10}\) &
\((3.21 \pm 1.76)\times10^{-9}\) &
\(1.83\sigma\) \\

\(A_2\) &
Transverse &
\((-1.38 \pm 0.57)\times10^{-13}\) &
\((-2.77 \pm 1.14)\times10^{-12}\) &
\(2.43\sigma\) \\

\(A_3\) &
Out of plane &
\((2.70 \pm 0.65)\times10^{-11}\) &
\((5.41 \pm 1.29)\times10^{-10}\) &
\(4.19\sigma\) \\
\bottomrule
\end{tabular}
\end{table}

The most recent normalized NGA components calculated in the paper \cite{Seligman2024DarkCometPopulations} are provided in Table \ref{tab:marsden_coefficients}, are they compatible with (to a first order of approximation) that generated by solar radiation pressure (SRP) on the solar panels of the spacecraft Phobos 1? The parameters adopted for Phobos 1 are summarised in Table \ref{tab:phobos1_parameters}. The acceleration induced by SRP on the panels of the Phobos 1 probe can be written as approximately:\\

\begin{equation}
    a = \frac{P_{\odot}C_{eff}}{M}
\end{equation}
 or alternatively:
 \begin{equation}
    C_{eff} = \frac{aM}{P_{\odot}}
\end{equation}
where $C_{eff}$ incorporates projected panel area and its reflectivities/absorptivity. Adopting the resultant NGA from Table \ref{tab:marsden_coefficients} of $\sim{3.3}$ $\times$ 10$^{-9}$ $\si{m.s^{-2}}$ implies a $C_{eff}$ $\sim{4.5}$ $\si{m^2}$. This effective area is on the same order as the estimate of Phobos 1’s solar panels, in fact it is a fraction 0.45 of it. The difference might be sensibly ascribed to the angle between the plane of the solar panels and the Sun-radial vector, and also the reflectivities
mentioned previously.

\begin{table}
\centering
\caption{Physical parameters adopted for the Phobos~1 solar-radiation-pressure model.}
\label{tab:phobos1_parameters}
\begin{tabular}{lc}
\toprule
Parameter & Value \\
\midrule
Phobos~1 mass, \(M\) & \(6220\ \mathrm{kg}\) \\
Solar-radiation pressure at 1 au, \(P_{\odot}\)
    & \(4.56\times10^{-6}\ \mathrm{N\,m^{-2}}\) \\
Approximate total solar-panel area, \(A_{\mathrm{panel}}\)
    & \(\sim 10\ \mathrm{m^2}\) \\
\bottomrule
\end{tabular}
\end{table}

\section{Fitting Different Spacecraft Geometries to Astrometry}

Here we take the spacecraft hypothesis and derive its consequences on the NGAs of 1998 KY$_{26}$, assuming they are caused principally by SRP on the spacecraft. To this end, the widely used orbit-determination software, $find\_orb$ \citep{Gray2022FindOrb}, which has the functionality to model NGAs and determine best fits for the model's parameters, was exploited and edited accordingly. It was downloaded on 22 August 2026 and various modifications made and stored locally. These modifications were to 5 files altogether and are summarised in the Appendix \ref{fo}, Table \ref{tab:findorb_modifications}.\\

The first task was to acquire the best gravity-only base line for the 260 observations of 1998 KY$_{26}$, using the standard unmodified version of $find\_orb$ in interactive mode, this is provided as the first data row of Table \ref{tab:ky26_provisional_model_fits}. This was subsequently used to compare against all models listed in this table, and indeed all of the fits were superior to this basic model in terms of RMS residual (refer last column).\\

We explored 3 'empirical' models, i.e. models which bear no direct physical relation to a real-world situation, but whose equations follow a general form one might expect from a spinning spacecraft's solar panels. Refer to Appendix \ref{app:empirical_nga_models} which describes these 3 models in detail. For a solar panel spinning about an axis parallel to its normal, then all SRP components perpendicular to the spin axis will cancel per spin revolution, leaving only a force along the spin axis itself. The equations adopted are inspired by the logic laid out in \cite{Hibberd2026Phobos1} for instance, and also respect the qualitative evolution demonstrated in Figure 1 of \cite{Farnocchia2025KY26}. In this empirical model there are 3 perpendicular and mutually independent components of surface area per unit mass, i.e. $\Gamma_R$, $\Gamma_N$ and $\Gamma_T$, for radial, normal to orbital plane and transverse respectively. In fact to properly account for a real solar panel, these would not be independent, however they do in fact express the expected harmonic structure of a real panel without respecting the interdependence of these $\Gamma$s. \\

Data rows 2, 3 and 4 show the results of these first order empirical models and demonstrate an appreciable reduction in residual from 0.894547 (gravity only) by $\sim{1} \%$ for the radial, a further $\sim{5.8} \%$ for the radial and normal, and then a further $\sim{5.4} \%$ for all the components including transverse. Thus the most important reproducible facets predicted by this simple empirical solar panel model of 1998 KY$_{26}$'s astrometric behaviour are those associated with the normal and transverse components of NGAs.\\

Since certain launch vehicle upper stages, such as Centaurs, can be well approximated by a cylinder, it would be instructive next to look at the consequence of a cylinder model for the NGAs of 1998 KY$_{26}$. To this end, the model described in Appendix \ref{cyl} was adopted based upon the theory expounded in \cite{Boulton1984CylindricalSRP}. First several pole raster scans of increasing resolution were performed on $(\lambda, \beta)$ across the entire sphere of possibilities, leading to a solution $( 321^{\circ}, 8^{\circ} )$ with minimum residual 0.799079, a $\sim{11} \%$ reduction w.r.t. the gravity only solution (refer Table \ref{tab:ky26_provisional_model_fits}). Note that the cylinder model is identical w.r.t. a 180$^{\circ}$ switch in principal axis, but is still an appreciable $\sim{21}^{\circ}$ off the photometric pole mentioned in Section \ref{sec:lc}.\\

For this cylinder model, the equation :

\begin{equation}
    \frac{h}{D} =\frac{\pi\Gamma_B}{4\Gamma_E}
\end{equation}

can be used to derive the ratio of thickness to diameter of the cylinder, whence from Table \ref{tab:ky26_provisional_model_fits}, we arrive at $\frac{h}{D}=1.37$. Thus the cylinder is 1.37 times taller along the principal axis than it is wide. This compares with the oblate spheroid solution of \cite{Farnocchia2025KY26} which had quite the reverse, i.e. a ratio of principal axis to equatorial axis of $\sim{0.7}$. These two parameters are near reciprocals and the difference might be reconciled in that the \cite{Farnocchia2025KY26} pole of $( 36^{\circ}, -44^{\circ} )$ is roughly 90$^{\circ}$ away from the cylinder pole, $( 321^{\circ}, 8^{\circ} )$, found here. \\

Next is the physically-justifiable vector solar panel model detailed in Appendix \ref{app:vector_panel_model}, summarised as the 7$^{th}$ row of results in Table \ref{tab:ky26_provisional_model_fits}. A coarse raster scan of the pole using the complete sweep of $( \lambda, \beta )$ was performed, followed by a localised medium scan, in turn followed by a fine resolution scan. The result was an optimal pole of $( 345^{\circ}, 4^{\circ} )$. The panel model adopted here is symmetrical about a change in sign of the assumed principal/spin axis, thus we have this and the photometric pole misaligned by $\sim{20}^{\circ}$. The RMS residual for this solution is 0.796897, representing a reduction of $\sim{11} \%$ compared to gravity only, almost matching the best empirical fit described above. \\

The vector panel model with the \cite{Farnocchia2025KY26} pole is slightly inferior to that using the photometric pole derived in Section \ref{sec:lc}.\\

It should be noted that the \cite{Marsden1973} model is superior in fidelity of fit to all other models investigated here, including the vector panel assumption, though the difference is a negligible $0.36\%$. The oblate spheroid model (the last entry in Table \ref{tab:ky26_provisional_model_fits}) is significantly worse than nearly all the solutions found here. However, the reader should be aware that the SRP-only model for an oblate spheroid will be lacking as far as transverse and normal NGA components are concerned, a difficulty which \cite{Farnocchia2025KY26} corrected by including a Yarkovsky-related treatment in addition to the straightforward SRP-only formulation.

\section{Discussion}\label{disc}

 Both photometry and astrometry were examined and first order approximation fits to the shape and SRP, respectively, were conducted to test the likelihood of the Phobos 1 hypothesis. Photometric comparisons indicate a strong agreement between the best fitting parameters of the Phobos 1 shape (where parameters include principal axis pole), against 8 light curves, though some underlying features at high solar phase are not present in the calculated light curves. This could possibly be due to the impact of simplifications (enforced by practical time constraints) on the fidelity of the model to the actual shape of Phobos 1 which has a complicated structure with the potential for odd harmonics. Furthermore, over several iterations of the Phobos 1 model, where new parameters were introduced for optimization (like solar panel width for example), their optimal values turned out to match those of the Phobos 1 probe, or at least to within the uncertainty achievable from the photographs of Phobos 1.\\

 An attempt to fit a solar panel response to SRP, reveals a noticeable reduction in observational residual of $\sim{11\%}$ compared to the gravity-only model, though the fit is still slightly inferior ($\sim{0.36\%}$) to the standard SRP model propounded by \cite{Marsden1973}. When one examines the results in terms of significance of the best-fit parameters for each model (Table \ref{tab:nga_parameter_significance}), there is clear evidence that the overwhelming advantage of the solar panel model (either empirical or the more physically-representative vector model) is the introduction of an acceleration component perpendicular to the orbital plane. In the empirical case this normal-to-orbit component is introduced directly by a 'normal parameter' which has a high $\sim{11\sigma}$ significance, whereas in the vector panel scenario this normal-to-orbit component is present indirectly through the $\Gamma_{panel,n}$ parameter which points normal to the panel, not necessarily to the orbital plane. Note the assumption of specular reflection for this panel-normal effect seems justified.\\

 Furthermore, when one examines the Sun-radial contribution in the realistic vector panel model, the $\Gamma_{pannel}$ significance is similarly high ($\sim{5\sigma}$), indicating a large absorptive component in these NGAs, as one would expect from solar panels.\\
 
 The separately calculated poles from photometry and astrometry are $(\lambda, \beta) = (151^{\circ}, +11^{\circ})$ and $( 165^{\circ}, -4^{\circ})$, respectively, where the latter assumes the solar panel model. The pointing angle error is 20$^{\circ}$, which is large, but it is proposed that this might be due to warped or twisted solar panels - or their struts - present on the probe but not modelled herein. \\

\begin{landscape}
\begin{table}[p]
\centering
\caption{Provisional comparison of nongravitational-force models fitted to
the 260-observation astrometric data set for 1998 KY$_{26}$.}
\label{tab:ky26_provisional_model_fits}
\begin{threeparttable}
\small
\setlength{\tabcolsep}{4pt}
\renewcommand{\arraystretch}{1.30}
\begin{tabularx}{\linewidth}{|p{3.1cm}|p{3.2cm}|X|p{3.4cm}|c|}
\hline
\textbf{Model} & \textbf{Pole or orbital phase}
& \textbf{Fitted physical parameters}
& \textbf{Parameter correlations}
& \textbf{Optical RMS (arcsec)}
\\ \hline

Gravity-only common reference
& --- & --- & --- & $\mathbf{0.894547}$
\\ \hline

Empirical radial only
& $f_0=\mathbf{228.0^\circ}$\newline
  {\footnotesize N.D. (scan)}
& $\Gamma_R=\mathbf{8.950\times10^{-4}}$\newline
  {\footnotesize $\sigma_{\Gamma_R}$ N.D. (scan)}
& --- & $\mathbf{0.885504}$
\\ \hline

Empirical radial $+$ normal
& $f_0=\mathbf{45^\circ}$\newline {\footnotesize N.D. (scan)}
& $\Gamma_R=\mathbf{1.70\times10^{-3}}$\newline
  {\footnotesize $\sigma_{\Gamma_R}$ N.D. (scan)}\newline
  $\Gamma_N=\mathbf{(-1.05746\pm0.09152)\times10^{-5}}$
& --- & $\mathbf{0.833962}$
\\ \hline

Empirical radial $+$ normal $+$ transverse
& $f_0=\mathbf{49.0^\circ}$\newline {\footnotesize N.D. (scan)}
& $\Gamma_R=\mathbf{1.1000\times10^{-3}}$\newline
  {\footnotesize $\sigma_{\Gamma_R}$ N.D. (scan)}\newline
  $\Gamma_N=\mathbf{(-8.5764\pm0.9000)\times10^{-6}}$\newline
  $\Gamma_T=\mathbf{(4.2084\pm1.3200)\times10^{-5}}$
& $\rho_{NT}=\mathbf{0.2197}$
& $\mathbf{0.796121}$
\\ \hline

Cylindrical SRP, common-state optimum
& $(\lambda,\beta)=\mathbf{(321^\circ,8^\circ)}$\newline
  {\footnotesize N.D. (scan); axis equivalent to $(141^\circ,-8^\circ)$}
& $\Gamma_B=\mathbf{(9.1070\pm2.1368)\times10^{-4}}$\newline
  $\Gamma_E=\mathbf{(5.2347\pm1.6640)\times10^{-4}}$
& $\rho_{BE}=\mathbf{0.9936}$
& $\mathbf{0.799079}$
\\ \hline

Cylindrical SRP, photometric axis
& $(\lambda,\beta)=\mathbf{(151^\circ,11^\circ)}$\newline
  {\footnotesize photometrically derived; fixed; axis equivalent to $(331^\circ,-11^\circ)$}
& $\Gamma_B=\mathbf{(2.5345\pm1.8575)\times10^{-4}}$\newline
  $\Gamma_E=\mathbf{(3.8161\pm1.4361)\times10^{-4}}$
& $\rho_{BE}=\mathbf{0.9912}$
& $\mathbf{0.817100}$
\\ \hline

Vector panel, common-state optimum
& $(\lambda,\beta)=\mathbf{(345^\circ,4^\circ)}$\newline
  {\footnotesize N.D. (scan)}
& $\Gamma_{\rm panel}=\mathbf{(1.4190\pm0.2795)\times10^{-3}}$\newline
  $\Gamma_{\rm panel,n}=\mathbf{(-1.1143\pm0.1018)\times10^{-4}}$
& $\rho_{\rm p,pn}=\mathbf{-0.0925}$
& $\mathbf{0.796897}$
\\ \hline

Vector panel, photometric pole\newline
{\footnotesize common reference state}
& $(\lambda,\beta)=\mathbf{(151^\circ,11^\circ)}$\newline
  {\footnotesize photometrically derived; fixed}
& $\Gamma_{\rm panel}=\mathbf{(7.1968\pm2.6699)\times10^{-4}}$\newline
  $\Gamma_{\rm panel,n}=\mathbf{(6.3777\pm0.6639)\times10^{-5}}$
& $\rho_{\rm p,pn}=\mathbf{-0.1109}$
& $\mathbf{0.812997}$
\\ \hline

Vector panel, Farnocchia pole\newline
{\footnotesize common reference state}
& $(\lambda,\beta)=\mathbf{(36^\circ,-44^\circ)}$\newline
  {\footnotesize adopted; not fitted}
& $\Gamma_{\rm panel}=\mathbf{1.34594\times10^{-3}}$\newline
  $\Gamma_{\rm panel,n}=\mathbf{5.35722\times10^{-5}}$\newline
  {\footnotesize dispersions N.D.}
& {\footnotesize N.D.}
& $\mathbf{0.818384}$
\\ \hline

Standard inverse-square Marsden
& {\footnotesize No pole or phase parameter}
& $A_1=\mathbf{(4.81\pm5.35)\times10^{-11}}$\newline
  $A_2=\mathbf{(-20.71\pm4.54)\times10^{-14}}$\newline
  $A_3=\mathbf{(11.8\pm1.3)\times10^{-12}}$\newline
  {\footnotesize in $\mathrm{au\,d^{-2}}$}
& {\footnotesize N.D.; full covariance matrix not retained}
& $\mathbf{0.794}$
\\ \hline

Analytical oblate-spheroid SRP\newline
{\footnotesize no Yarkovsky term}
& $(\lambda,\beta)=\mathbf{(36^\circ,-44^\circ)}$\newline
  {\footnotesize fixed; not fitted}
& $\rho=\mathbf{615.52\ {\rm kg\,m^{-3}}}$\newline
  {\footnotesize $\sigma_\rho$ N.D.}\newline
  $A/M=\mathbf{3.12031\times10^{-4}}$\newline
  {\footnotesize derived from fitted $\rho$}
& {\footnotesize N/A: one fitted physical parameter}
& $\mathbf{0.818501}$
\\ \hline

\end{tabularx}
\begin{tablenotes}[flushleft]\scriptsize
\item[] Unless stated otherwise, $\Gamma$ is in
$\mathrm{m^2\,kg^{-1}}$ and quoted uncertainties are formal
$1\sigma$ dispersions conditional on fixed scanned quantities.
N.D. denotes not determined; N/A denotes not applicable.
Marsden coefficients are in $\mathrm{au\,d^{-2}}$ and use
$g(r)=r^{-2}$. Pole coordinates and $f_0$ values selected by raster
search have no formal covariance uncertainties. A cylindrical pole denotes
an undirected axis, so $(\lambda,\beta)$ is equivalent to
$(\lambda+180^\circ,-\beta)$ modulo $360^\circ$. All common-state fits
used 260 observations and a gravity-reference RMS of $0.894547''$.
\end{tablenotes}
\end{threeparttable}
\end{table}
\end{landscape}

\begin{table}[htbp]
\centering
\caption{Formal significance of fitted nongravitational parameters.
Significances are conditional on the adopted or raster-selected phase
or pole.}
\label{tab:nga_parameter_significance}
\small
\setlength{\tabcolsep}{5pt}
\renewcommand{\arraystretch}{1.20}
\begin{tabular}{|l|c|c|c|}
\hline
\textbf{Model} &
\textbf{Parameter} &
\textbf{Estimate $\boldsymbol{\pm\,1\sigma}$} &
\textbf{Significance}
\\ \hline

Empirical radial $+$ normal
& $\Gamma_N$
& $(-1.05746 \pm 0.09152)\times10^{-5}$
& $11.55\sigma$
\\ \hline

Empirical radial $+$ normal $+$ transverse
& $\Gamma_N$
& $(-8.5764 \pm 0.9000)\times10^{-6}$
& $9.53\sigma$
\\

& $\Gamma_T$
& $(4.2084 \pm 1.3200)\times10^{-5}$
& $3.19\sigma$
\\ \hline

Cylindrical SRP, raster optimum
& $\Gamma_B$
& $(9.1070 \pm 2.1368)\times10^{-4}$
& $4.26\sigma$
\\

& $\Gamma_E$
& $(5.2347 \pm 1.6640)\times10^{-4}$
& $3.15\sigma$
\\ \hline

Cylindrical SRP, photometric pole
& $\Gamma_B$
& $(2.5345 \pm 1.8575)\times10^{-4}$
& $1.36\sigma$
\\

& $\Gamma_E$
& $(3.8161 \pm 1.4361)\times10^{-4}$
& $2.66\sigma$
\\ \hline

Vector-panel SRP, raster optimum
& $\Gamma_{\rm panel}$
& $(1.4190 \pm 0.2795)\times10^{-3}$
& $5.08\sigma$
\\

& $\Gamma_{\rm panel,n}$
& $(-1.1143 \pm 0.1018)\times10^{-4}$
& $10.95\sigma$
\\ \hline

Vector-panel SRP, photometric pole
& $\Gamma_{\rm panel}$
& $(7.1968 \pm 2.6699)\times10^{-4}$
& $2.70\sigma$
\\

& $\Gamma_{\rm panel,n}$
& $(6.3777 \pm 0.6639)\times10^{-5}$
& $9.61\sigma$
\\ \hline

Marsden $A_1+A_2+A_3$
& $A_1$
& $(4.81 \pm 5.35)\times10^{-11}\ \mathrm{au\,d^{-2}}$
& $0.90\sigma$
\\

& $A_2$
& $(-20.71 \pm 4.54)\times10^{-14}\ \mathrm{au\,d^{-2}}$
& $4.56\sigma$
\\

& $A_3$
& $(11.8 \pm 1.3)\times10^{-12}\ \mathrm{au\,d^{-2}}$
& $9.08\sigma$
\\ \hline

Analytical oblate-spheroid SRP
& $\rho$
& $(527 \pm 49)\ \mathrm{kg\,m^{-3}}$
& $10.76\sigma$
\\ \hline
\end{tabular}

\vspace{2mm}
\parbox{0.96\linewidth}{\footnotesize
All $\Gamma$ values are in $\mathrm{m^2\,kg^{-1}}$.
The radial coefficients $\Gamma_R$, phase $f_0$, and pole coordinates
selected by raster scanning are excluded because no covariance
dispersions were calculated for them. These are formal local
significances and do not incorporate uncertainty in the selected phase,
pole, force law, or astrometric systematics.}
\end{table}

\section{Conclusion}\label{conc}

In this paper the feasibility of the hypothesis that the dark comet 1998 KY$_{26}$ could be the Soviet Phobos 1 probe was investigated. The combined orbital, astrometric, radar and photometric evidence makes Phobos 1 a strongly plausible but presently unproven identification for 1998 KY$_{26}$ due to the complex nongravitational accelerations and remaining residual features in observed photometry at high-phase-angles not entirely reproduced by the Phobos 1 model. These are most likely explained by simplifications in this derived model. All shall be revealed when the Hayabusa2$\sharp$ spacecraft intercepts the object in July 2031.

\section{Acknowledgements}
AI was used in this investigation.

\bibliographystyle{cas-model2-names}

\bibliography{Phobos1_Hypothesis}


\clearpage
\appendix

\section{Photometric Pole Search for the Phobos~1 Model}
\label{app:pole_search}

This appendix describes the numerical photometric model used to search for the spin-axis orientation of 1998~KY$_{26}$ under the hypothesis that the object has the approximate geometry of the Phobos~1 spacecraft. The final model contains two solar-panel wings, bright inner-edge panel stripes, a cylindrical payload bus, two mutually orthogonal radiator plates, a one-sided high-gain antenna, four upper spherical tanks, a sampled toroidal structural component, and a simplified autonomous propulsion unit (ADU). Solar shadowing and observer-side occultation by these components were evaluated explicitly.

The rotation period was fixed at
\begin{equation}
P_{\rm rot}=5.3516~\mathrm{min}.
\label{app:rotation_period}
\end{equation}

The spacecraft geometry was developed through a sequence of diagnostic scans. The solar-panel breadth, radiator brightness, stripe brightness and radiator inclination were investigated before the final pole refinement. Consequently, the final pole search did not freely alter the principal dimensions of the spacecraft at every trial pole.

\subsection{Photometric data and statistical weights}
\label{app:photometric_data}

The source photometry comprised 954 measurements divided among 22 independently calibrated light-curve segments and eight observing groups. Each source segment retained its own additive magnitude zero point, while each observing group retained an independent rotational-phase offset.

For a measurement at time $t_i$, the initial rotational phase was
\begin{equation}
\phi_i=\operatorname{mod}\left(\frac{t_i-t_{{\rm ref},g}}{P_{\rm rot}},1\right),
\label{app:initial_phase}
\end{equation}
where $t_{{\mathrm{ref}},g}$ is the reference epoch for observing group $g$.

The source table did not provide homogeneous per-measurement uncertainties. An empirical scatter was therefore estimated separately for each segment $s$ from successive magnitude differences (see also \cite{Czesla2018}):
\begin{equation}
\sigma_s=\frac{1.4826}{\sqrt{2}}\operatorname{median}\left(\left|\Delta m_i-\operatorname{median}(\Delta m_i)\right|\right),
\qquad \Delta m_i=m_{i+1}-m_i.
\label{app:segment_sigma}
\end{equation}
For numerical stability, $0.03~\mathrm{mag}\leq\sigma_s\leq0.20~\mathrm{mag}$. These quantities supplied relative statistical weights and should not be interpreted as independently calibrated measurement uncertainties.

\subsection{Spin-axis coordinates and rotational geometry}
\label{app:pole_coordinates}

The directed spin axis was expressed by its ecliptic J2000 longitude and latitude, $\lambda$ and $\beta$:
\begin{equation}
\hat{\boldsymbol{p}}=
\begin{pmatrix}
\cos\beta\cos\lambda\\
\cos\beta\sin\lambda\\
\sin\beta
\end{pmatrix}.
\label{app:pole_vector}
\end{equation}
A right-handed body basis was constructed using
\begin{equation}
\hat{\boldsymbol{x}}_{\rm b}=\frac{\hat{\boldsymbol{k}}\times\hat{\boldsymbol{p}}}{|\hat{\boldsymbol{k}}\times\hat{\boldsymbol{p}}|},
\qquad
\hat{\boldsymbol{y}}_{\rm b}=\hat{\boldsymbol{p}}\times\hat{\boldsymbol{x}}_{\rm b},
\label{app:body_basis}
\end{equation}
where $\hat{\boldsymbol{k}}$ was replaced by a non-parallel reference vector near an ecliptic pole.

At rotational angle $\psi=2\pi\phi$, the spacecraft was held fixed and the Sun and observer vectors were rotated about body $z$:
\begin{equation}
\begin{split}
s_x(\psi)&=\cos\psi\,s_x+\sin\psi\,s_y,\\
s_y(\psi)&=-\sin\psi\,s_x+\cos\psi\,s_y,\\
s_z(\psi)&=s_z,
\end{split}
\label{app:rotated_sun}
\end{equation}
with the same transformation applied to the observer vector. Mean Sun and observer directions were used within each source segment, while individual measurement times were retained when calculating rotational phase.

\subsection{Adopted spacecraft geometry}
\label{app:spacecraft_geometry}

The principal dimensions and treatments used in the final pole search are listed in Table~\ref{tab:phobos_photometric_model}. Dimensions inferred from photographs are necessarily approximate.

\begin{table}[htbp]
\centering
\caption{Adopted parameters of the final Phobos~1 photometric model.}
\label{tab:phobos_photometric_model}
\small
\begin{tabular}{@{}p{0.29\linewidth}p{0.64\linewidth}@{}}
\hline
\textbf{Component} & \textbf{Adopted value or treatment}\\
\hline
Rotation & Period $5.3516~\mathrm{min}$.\\
Solar panels & $R_{\mathrm{in}}=2.40~\mathrm{m}$, $R_{\mathrm{out}}=5.00~\mathrm{m}$, active breadth $W=2.50~\mathrm{m}$, and combined modelled area $13.0~\mathrm{m^2}$.\\
Inner stripes & Depth $0.60~\mathrm{m}$; one complete stripe and one stripe with a $0.75~\mathrm{m}$ central interruption.\\
Payload bus & Right circular cylinder of radius $1.14~\mathrm{m}$ and height $2.189801~\mathrm{m}$.\\
Radiators & Two $1.60\times0.55~\mathrm{m}$ plates at centre radius $1.65~\mathrm{m}$, clocks $0^\circ$ and $-90^\circ$, with tangential long axes. Inclinations $0^\circ$ and $35^\circ$.\\
High-gain antenna & Flat disk of radius $0.60~\mathrm{m}$, cant $70^\circ$, centre on the spin axis, and final clock $195^\circ$.\\
Upper spherical tanks & Four spheres of radius $0.30~\mathrm{m}$, ring radius $1.45~\mathrm{m}$, $z=+0.62~\mathrm{m}$, and initial clock $45^\circ$.\\
Toroidal structure & 20 spherical blockers; major radius $1.13~\mathrm{m}$, minor radius $0.27~\mathrm{m}$, and $z=0$.\\
Large ADU tanks & Four spheres of radius $0.90~\mathrm{m}$, ring radius $1.45~\mathrm{m}$, and $z=-1.65~\mathrm{m}$.\\
Small ADU tanks & Four spheres of radius $0.34~\mathrm{m}$, ring radius $0.78~\mathrm{m}$, $z=-1.65~\mathrm{m}$, and clock offset $45^\circ$.\\
Engine/support assembly & Central spherical blocker of radius $0.38~\mathrm{m}$ at $z=-2.25~\mathrm{m}$.\\
\hline
\end{tabular}
\end{table}

The panels were treated as coplanar, opaque and infinitesimally thin rectangles in the body $z=0$ plane. Direct panel reflection was present only when the Sun and observer lay on the same side of this plane. Before masking, the panel contribution was proportional to
\begin{equation}
F_{{\rm panel},0}=A_{\rm panel}|s_z||o_z|.
\label{app:unblocked_panel_flux}
\end{equation}
The panel surfaces were rastered, and samples whose rays towards the Sun or observer intersected another component were removed. If $f_{\mathrm{visible}}(\psi)$ is the fraction simultaneously illuminated and visible, then
\begin{equation}
F_{\rm panel}(\psi)=
\begin{cases}
F_{{\rm panel},0}f_{\rm visible}(\psi),&s_zo_z>0,\\
0,&s_zo_z\leq0.
\end{cases}
\label{app:panel_flux}
\end{equation}

The $0.60~\mathrm{m}$-deep cells forming the white inner-edge stripes were retained in the ordinary panel template and also stored as a separate stripe template. The $+x$ stripe was complete. The corresponding $-x$ stripe was interrupted over its central $0.75~\mathrm{m}$. This avoided removing the underlying panel where the stripe enhancement was absent.

The bus brightness was calculated by integrating Lambertian reflection over the barrel and upper circular end of a right circular cylinder. For outward surface normal $\hat{\boldsymbol{n}}$, each element contributed in proportion to
\begin{equation}
\max(0,\hat{\boldsymbol{n}}\!\cdot\!\hat{\boldsymbol{s}})\max(0,\hat{\boldsymbol{n}}\!\cdot\!\hat{\boldsymbol{o}})\,dA.
\label{app:lambert_surface_element}
\end{equation}
The barrel integral was evaluated using 360 azimuthal samples. The upper end was integrated analytically; the lower circular end was assigned zero direct brightness because it is largely obscured by the toroidal chassis and ADU in the adopted cruise configuration.

The high-gain antenna was represented by an opaque circular raster. Its front face had the fitted coefficient $q_{\mathrm{ant}}$. In the implemented final model, the rear face used the bus-like coefficient $q_{\mathrm{bus}}$, rather than being perfectly dark. A face contributed only when both Sun and observer were on that side of the disk. Panel and radiator intersections were included in its shadow and occultation masks.

The two radiator plates were separated by $90^\circ$ in body azimuth. Their $1.60~\mathrm{m}$ axes were tangential and therefore perpendicular to their respective radius vectors from the spacecraft $z$-axis. The short axes were radial when the plate inclination was zero. For the inclined plate, the inner long edge remained fixed and the outer edge tilted downward by $35^\circ$. As implemented in the final pole scan, both faces used the same diffuse coefficient. Each plate was rastered and independently tested for shadowing and occultation.

The four large and four small ADU tanks, the toroidal samples and the central engine proxy entered the final pole scan principally as opaque shadowing and occulting bodies. Their unresolved, approximately rotation-independent direct brightness was absorbed into $q_{\mathrm{bus}}$. The four upper spherical tanks additionally supplied an optional sampled specular template. Their spherical surfaces were represented by 362 approximately equal-area directions and a normalised Blinn--Phong lobe with fixed exponent $m_{\mathrm{metal}}=30$. The amplitude of this component, $\eta_{\mathrm{metal}}$, was fitted rather than assumed.

\subsection{Model flux and fitted coefficients}
\label{app:total_flux}

The relative model flux was
\begin{equation}
\begin{split}
F_i={}&F_{{\rm panel},i}+(\eta_{\rm stripe}-1)F_{{\rm stripe},i}
+q_{\rm bus}\left(F_{{\rm bus},i}+F_{{\rm ant,rear},i}\right)\\
&+q_{\rm ant}F_{{\rm ant,front},i}
+\eta_{\rm board}q_{\rm bus}F_{{\rm board},i}
+\eta_{\rm metal}F_{{\rm metal},i}.
\end{split}
\label{app:total_flux_equation}
\end{equation}
Thus $q_{\mathrm{board}}=\eta_{\mathrm{board}}q_{\mathrm{bus}}$. All brightness quantities are effective coefficients relative to the unit solar-panel template; they are not measured material albedos. The subtraction of unity from $\eta_{\mathrm{stripe}}$ prevents the stripe cells from being counted twice.

The raw and calibrated model magnitudes were
\begin{equation}
m_{i,{\rm raw}}=-2.5\log_{10}F_i,
\qquad m_{i,{\rm model}}=m_{i,{\rm raw}}+Z_{s(i)}.
\label{app:model_magnitude}
\end{equation}
For fixed physical parameters and phases, the optimum zero point for segment $s$ was calculated analytically:
\begin{equation}
Z_s=\frac{\sum_{i\in s}w_i(m_{i,{\rm obs}}-m_{i,{\rm raw}})}{\sum_{i\in s}w_i},
\qquad w_i=\sigma_s^{-2}.
\label{app:zero_point}
\end{equation}

The principal objective was
\begin{equation}
\chi^2=\sum_i\left[\frac{m_{i,{\rm obs}}-m_{i,{\rm model}}}{\sigma_{s(i)}}\right]^2,
\label{app:chi_square}
\end{equation}
and the unweighted pointwise RMS was retained as a diagnostic:
\begin{equation}
{\rm RMS}=\left[\frac{1}{N}\sum_i(m_{i,{\rm obs}}-m_{i,{\rm model}})^2\right]^{1/2}.
\label{app:photometric_rms}
\end{equation}

\subsection{Staged geometrical and pole optimisation}
\label{app:optimization}

The analysis was conducted hierarchically. An initial all-sky pole raster was followed by targeted comparisons of the leading pole regions. Separate scans then investigated the active panel breadth and inner radius, one- and two-radiator configurations, rotation sense, stripe brightness and radiator inclination. This was followed by all-group pole-cap scans and a final local two-stage pole grid.

The active panel breadth was not selected solely from the photographs. Photometric scans showed that a breadth close to
\begin{equation}
W=2.50~\mathrm{m}
\label{app:preferred_panel_width}
\end{equation}
was required to reproduce the widths and rotational locations of important light-curve features, particularly at the $82^\circ$ solar phase angle. This agrees with the breadth estimated independently from the available photographs of Phobos~1. The radiator-inclination scan independently favoured approximately $35^\circ$, also consistent with the visibly inclined radiator in the photographs.

During the final all-group pole search, the antenna clock, panel geometry, radiator inclination and stripe enhancement were held at
\begin{equation}
\gamma_{\rm ant}=195^\circ,
\quad W=2.50~\mathrm{m},
\quad R_{\rm in}=2.40~\mathrm{m},
\quad \delta_{\rm board}=35^\circ,
\quad \eta_{\rm stripe}=8.
\label{app:final_fixed_geometry}
\end{equation}

At every trial pole, the common parameters $q_{\mathrm{bus}}$, $q_{\mathrm{ant}}$, $\eta_{\mathrm{board}}$ and $\eta_{\mathrm{metal}}$ were profiled against the combined $\chi^2$ of all eight observing groups. Their bounds were
\begin{equation}
10^{-5}\leq q_{\rm bus}\leq12,
\quad 10^{-6}\leq q_{\rm ant}\leq5,
\quad 0\leq\eta_{\rm board}\leq12,
\quad 0\leq\eta_{\rm metal}\leq20.
\label{app:coefficient_bounds}
\end{equation}
Bounds were enforced using logistic transformations, and the transformed parameters were optimized by the Nelder--Mead method implemented in \texttt{fminsearch}. Multiple initial simplexes were used. For every proposed set of common coefficients, the eight group phase offsets were independently profiled using a uniform phase scan followed by bounded scalar minimisation.

The final rapid refinement used 72 rotational template samples and 36 initial phase samples. Its first stage evaluated a $5\times5$ grid with $0.5^\circ$ spacing around $(150^\circ,+12^\circ)$. Its second stage evaluated a $3\times3$ grid with $0.25^\circ$ spacing centred automatically on the first-stage winner. Every completed pole was checkpointed to CSV and MAT files.

\subsection{Preferred solution}
\label{app:preferred_solution}

The final two-stage scan reproduced the preferred solution at
\begin{equation}
\boxed{(\lambda,\beta)=(151^\circ,+11^\circ)}.
\label{app:preferred_pole}
\end{equation}
The same pole and objective value were recovered in independent local scans. The adopted pole uncertainty is

\begin{equation}
      (\lambda,\beta)=({151^{+1}_{-2}}^{\circ},{+11^{+12}_{- 3}}^{\circ}),  
\label{app:pole_uncertainty}
\end{equation}

At the preferred grid point, the fitted coefficients and fit statistics were
\begin{equation}
q_{\rm bus}=3.1203,
\qquad q_{\rm ant}=0.47222,
\qquad \eta_{\rm board}=1.3518,
\qquad \eta_{\rm metal}=1.25\times10^{-7},
\label{app:preferred_coefficients}
\end{equation}
\begin{equation}
\chi^2_{\rm total}=934.2274,
\qquad {\rm RMS}=0.138753~\mathrm{mag},
\label{app:preferred_statistics}
\end{equation}
with
\begin{equation}
\chi^2_{82}=106.0412,
\qquad \chi^2_{101}=158.5190.
\label{app:high_phase_statistics}
\end{equation}

The fitted metallic-sphere coefficient is effectively zero. Thus an explicit specular contribution from the upper spherical tanks was not required by the all-group fit, although those spheres remained present in the obstruction geometry. In contrast, the radiator coefficient remained finite and stable.

The evidential significance of the fit does not arise solely from the number of adjustable photometric coefficients. Several parameters associated with identifiable spacecraft components converged towards values independently expected from the photographs. In particular, the light curves required an active panel breadth close to $2.5~\mathrm{m}$, and the preferred radiator inclination was approximately $35^\circ$. These quantities control the locations, durations and shapes of shadowing and occultation features rather than merely rescaling the total flux. Their agreement with the photographed Phobos~1 configuration therefore constitutes positive supporting evidence for the spacecraft-identification hypothesis.

\subsection{Principal limitations}
\label{app:model_limitations}

\begin{enumerate}
\item Component dimensions and locations were estimated from small photographs rather than complete engineering drawings or a CAD model.
\item The panels and radiators were treated as infinitesimally thin rectangles.
\item As implemented in the final pole scan, both faces of each radiator had the same diffuse response; detailed struts and attachments were omitted.
\item The antenna was represented by a flat disk, with its rear face assigned the bus-like brightness coefficient.
\item The large and small ADU tanks entered principally through obstruction; their phase-dependent diffuse brightness was not evaluated as a separate fitted template in the final pole scan.
\item The toroidal chassis and engine/support assembly were represented by spherical blocking proxies.
\item Only the optional upper-tank term included an explicit specular law; detailed wavelength-dependent or mixed diffuse--specular scattering was not modelled.
\item Mean Sun and observer vectors were used within each source segment.
\item Principal-axis rotation and a fixed rotation period were assumed.
\item Empirical segment dispersions were used because homogeneous pointwise uncertainties were unavailable.
\item Independent group phase offsets prevent maintenance of an absolute rotation count between observing epochs.
\item The quoted pole uncertainty is model-dependent and is not a formal covariance uncertainty.
\item Thermal emission was neglected.
\end{enumerate}

\clearpage
\section{Alterations made to $find\_orb$}\label{fo}
Refer to Table \ref{tab:findorb_modifications}. 
\begin{table*}
\centering
\caption{Broad summary of the principal changes made to the standard
\textsc{Find\_Orb} software for the analysis of the nongravitational
acceleration of 1998 KY$_{26}$.}
\label{tab:findorb_modifications}
\begin{tabular}{p{0.15\textwidth} p{0.43\textwidth} p{0.34\textwidth}}
\toprule
Source file & Broad modification & Purpose \\
\midrule

\texttt{mpc\_obs.h}
&
Introduced a general physical-model configuration structure containing
model type, geometry, orientation, optical properties and
area-to-mass coefficients. Added parameter identifiers and fitting flags
for oblate-spheroid, cylindrical and solar-panel models, and increased
the permitted number of fitted orbital and physical parameters.
&
Provides the common data structures through which the additional
nongravitational-force models and their fitted parameters are communicated
between the orbit determination, force integration and output routines.
\\[0.6em]

\texttt{runge.cpp}
&
Extended the equations of motion with additional solar-radiation-pressure
models. These include nonspherical oblate and cylindrical representations,
an empirical rotating-panel model, and vector-panel force calculations.
The empirical model permits independently scaled radial, normal and
transverse components with prescribed orbital-phase dependence.
&
Calculates the additional nongravitational acceleration at every
integration step and adds it to the standard gravitational equations of
motion.
\\[0.6em]

\texttt{elem\_out.cpp}
&
Added the high-level experimental control for the KY26 analysis:
physical-model selection, parameter initialization and mapping, raster
and profile scans, multistart fitting, matched comparisons between nested
force models, convergence protection, solution rollback and detailed
reporting of fitted coefficients and residuals.
&
Controls the sequence of orbit fits and searches the physical-parameter
space while ensuring that competing acceleration models are compared
using the same observations, weighting and initial orbital state.
\\[0.6em]

\texttt{orb\_func.cpp}
&
Extended the differential-correction and covariance machinery to treat
the new physical quantities as additional solve-for parameters. Added
parameter-specific numerical-differentiation increments, scaling and
trust-region treatment.
&
Allows the six-component orbital state and selected physical-force
parameters to be optimized together and permits formal parameter
uncertainties and correlations to be calculated.
\\[0.6em]

\texttt{mpc\_obs.cpp}
&
The standard observation-handling implementation was retained without a
specific alteration for the final transverse model, but was included in
the controlled five-file source package used to build the working
executable.
&
Preserves compatibility with the standard optical and radar observation
processing and provides a reproducible source set for the modified build.
\\

\bottomrule
\end{tabular}
\end{table*}
\clearpage
\section{Models Adopted for the NGAs in the Modified $find\_orb$}
\subsection{Cylindrical Model}\label{cyl}

From \cite{Boulton1984CylindricalSRP}, we have:

\begin{equation}
    \mathbf{F} = -\frac{P_{\odot}}{r^2}\left(F_1(p_1)\hat{\mathbf{r}}_{\odot}+F_2(p_1)\hat{\mathbf{p}}\right)
\end{equation}
where:
\begin{equation}
    p_1=\hat{\mathbf{r}}_{\odot}\cdot\hat{\mathbf{p}}
\end{equation}
and:
\begin{equation}
    F_1=\pi R^2(1-\rho_E)|p_1| + Rh\left[\frac{\pi}{2}\rho_s+2\left(1+\frac{\rho_s}{3}\right)(1-p_1^2)^{\frac{1}{2}}\right]
\end{equation}
\begin{equation}
    F_2=p_1\left[\pi \left(\delta_E R^2 - \delta_s \frac{Rh}{2}\right) +2\pi R^2\rho_E|p_1|-\frac{8}{3}Rh\rho_s(1-p_1^2)^{\frac{1}{2}}\right]
\end{equation}
In the above $\hat{\mathbf{r}}_{\odot}$ is the unit vector towards the Sun as opposed to $find\_orb$'s anti-solar convention, so the equations were quite straightforwardly modified accordingly.  $\hat{\mathbf{p}}$ represents the unit vector along the cylinder's central longitudinal axis. With the four cylinder reflectivities corresponding to $\rho_s$ and $\delta_s$, the specular and diffuse reflectivities of the barrel respectively, and those of the flat ends of the cylinder, $\rho_E$ and $\delta_E$, the above equations offer plenty of opportunity for unwanted degeneracy in the equations, thus a simpler form of the above equations must be derived with fewer parameters to be fitted.\\

Thus the assumption that the specular reflectivities are zero, and the diffuse reflectivities are the same at the barrel and the ends ($\delta$ was hardwired in the code as 0.29) yields:

\begin{equation}
    a_1 = \Gamma_E|p_1|+\Gamma_B(1-p_1^2)^{\frac{1}{2}}
\end{equation}
\begin{equation}
    a_2 = p_1\delta\left(\Gamma_E-\frac{\pi}{4}\Gamma_B\right)
\end{equation}

where $a_1$ and $a_2$ are now accelerations, with $\Gamma_E$ and $\Gamma_B$ as the areas per unit mass of the end and of the barrel longitudinal cross-section respectively, thus:
\begin{equation}
    \Gamma_E = \frac{\pi R^2}{M}
\end{equation}
\begin{equation}
    \Gamma_B = \frac{2Rh}{M}
\end{equation}

and:
\begin{equation}
    \mathbf{a} = -P\left(a_1(p_1)\hat{\mathbf{r}}_{\odot}+a_2(p_1)\hat{\mathbf{p}}\right)
\end{equation}

With the equations expressed in this way, there are now only 4 unknowns, the two Euler angles of the pole and the parameters $\Gamma_E$ and $\Gamma_B$.

\subsection{Empirical Nongravitational-Acceleration Models}
\label{app:empirical_nga_models}

This appendix defines the three nested empirical nongravitational-acceleration
models used in the astrometric analysis: radial (\(R\)), radial--normal
(\(RN\)), and radial--normal--transverse (\(RNT\)). Their angular dependence
was chosen to represent the lowest-order symmetries expected from a
two-sided, panel-like object. These models describe the acceleration required
by the astrometry but do not constitute complete physical models of the
spacecraft.

\subsubsection{Osculating orbital frame}

Let the heliocentric position and velocity of the object be
\(\mathbf{r}\) and \(\mathbf{v}\), respectively. The radial unit vector is

\begin{equation}
    \hat{\mathbf{R}}
    =
    \frac{\mathbf{r}}{r},
    \qquad
    r=|\mathbf{r}|.
    \label{eq:empirical_radial_unit}
\end{equation}

The specific orbital-angular-momentum vector is

\begin{equation}
    \mathbf{h}
    =
    \mathbf{r}\times\mathbf{v},
    \label{eq:empirical_angular_momentum}
\end{equation}

and the orbital-normal unit vector is

\begin{equation}
    \hat{\mathbf{N}}
    =
    \frac{\mathbf{h}}{|\mathbf{h}|}.
    \label{eq:empirical_normal_unit}
\end{equation}

The transverse unit vector, positive in the direction of orbital motion, is

\begin{equation}
    \hat{\mathbf{T}}
    =
    \hat{\mathbf{N}}\times\hat{\mathbf{R}}.
    \label{eq:empirical_transverse_unit}
\end{equation}

The three vectors form a right-handed orthonormal triad satisfying

\begin{equation}
    \hat{\mathbf{R}}\times\hat{\mathbf{T}}
    =
    \hat{\mathbf{N}}.
\end{equation}

\subsubsection{Calculation of true anomaly}

The eccentricity vector is

\begin{equation}
    \mathbf{e}
    =
    \frac{\mathbf{v}\times\mathbf{h}}{\mu_{\odot}}
    -
    \hat{\mathbf{R}},
    \label{eq:empirical_eccentricity_vector}
\end{equation}

where \(\mu_{\odot}=GM_{\odot}\). Defining

\begin{equation}
    \hat{\mathbf{e}}
    =
    \frac{\mathbf{e}}{|\mathbf{e}|}
\end{equation}

and

\begin{equation}
    \hat{\mathbf{q}}
    =
    \hat{\mathbf{N}}\times\hat{\mathbf{e}},
\end{equation}

the cosine and sine of the true anomaly are obtained from

\begin{align}
    \cos f
    &=
    \hat{\mathbf{e}}\cdot\hat{\mathbf{R}},
    \\
    \sin f
    &=
    \hat{\mathbf{q}}\cdot\hat{\mathbf{R}}.
\end{align}

The true anomaly used by the numerical implementation is therefore

\begin{equation}
    f
    =
    \operatorname{atan2}
    \left(
        \hat{\mathbf{q}}\cdot\hat{\mathbf{R}},
        \hat{\mathbf{e}}\cdot\hat{\mathbf{R}}
    \right).
    \label{eq:empirical_true_anomaly}
\end{equation}

A phase offset \(f_{0}\) locates the empirical acceleration pattern relative
to perihelion. The phase entering the force law is

\begin{equation}
    \psi=f-f_{0}.
    \label{eq:empirical_phase}
\end{equation}

The parameter \(f_{0}\) is an orbital-phase offset and must not be interpreted
as a pole longitude.

\subsubsection{Radial projected-area term}

For a two-sided flat surface, the projected illuminated area is proportional
to the absolute cosine of the incidence angle. The adopted radial modulation
is consequently

\begin{equation}
    g_R(\psi)
    =
    |\cos\psi|.
    \label{eq:empirical_radial_modulation}
\end{equation}

This function possesses twofold symmetry because

\begin{equation}
    |\cos(\psi+\pi)|
    =
    |\cos\psi|.
\end{equation}

Let \(P_{\odot}\) denote the solar-radiation pressure at \(1\,\mathrm{au}\),
with the heliocentric distance \(r\) expressed in astronomical units. The
radial acceleration is

\begin{equation}
    \mathbf{a}_{R}
    =
    \frac{P_{\odot}}{r^{2}}
    \Gamma_R
    |\cos(f-f_{0})|
    \hat{\mathbf{R}},
    \label{eq:empirical_radial_acceleration}
\end{equation}

where \(\Gamma_R\) is an effective area-to-mass coefficient. For
\(\Gamma_R>0\), the radial acceleration is always directed away from the Sun.

\subsubsection{Radial model}

The radial-only model is

\begin{equation}
    \boxed{
    \mathbf{a}_{\mathrm{NG}}^{(R)}
    =
    \frac{P_{\odot}}{r^{2}}
    \Gamma_R
    |\cos(f-f_{0})|
    \hat{\mathbf{R}}
    }.
    \label{eq:empirical_R_model}
\end{equation}

Its nongravitational parameters are \(f_{0}\) and \(\Gamma_R\). In the
numerical implementation, these quantities were raster-scanned, with the
six-component orbital state improved at each sampled point.

\subsubsection{Normal harmonic}

The normal component was required to reverse direction at diametrically
opposite orbital phases. The lowest-order harmonic with this property is

\begin{equation}
    g_N(\psi)
    =
    \cos\psi,
    \label{eq:empirical_normal_modulation}
\end{equation}

because

\begin{equation}
    \cos(\psi+\pi)
    =
    -\cos\psi.
\end{equation}

The normal acceleration is therefore

\begin{equation}
    \mathbf{a}_{N}
    =
    \frac{P_{\odot}}{r^{2}}
    \Gamma_N
    \cos(f-f_{0})
    \hat{\mathbf{N}}.
    \label{eq:empirical_normal_acceleration}
\end{equation}

For positive \(\Gamma_N\), the acceleration is directed along
\(+\hat{\mathbf{N}}\) when \(\cos(f-f_{0})>0\), and along
\(-\hat{\mathbf{N}}\) when \(\cos(f-f_{0})<0\). A negative fitted value of
\(\Gamma_N\) reverses this convention.

\subsubsection{Radial--normal model}

Combining Equations~\eqref{eq:empirical_radial_acceleration} and
\eqref{eq:empirical_normal_acceleration} gives

\begin{equation}
    \boxed{
    \mathbf{a}_{\mathrm{NG}}^{(RN)}
    =
    \frac{P_{\odot}}{r^{2}}
    \left[
        \Gamma_R
        |\cos(f-f_{0})|
        \hat{\mathbf{R}}
        +
        \Gamma_N
        \cos(f-f_{0})
        \hat{\mathbf{N}}
    \right]
    }.
    \label{eq:empirical_RN_model}
\end{equation}

The radial-only model is recovered by setting

\begin{equation}
    \Gamma_N=0.
\end{equation}

The quantities \(f_{0}\) and \(\Gamma_R\) were raster-scanned in the final
implementation. At each sampled point, \(\Gamma_N\) and the six-component
orbital state were determined by differential correction.

\subsubsection{Transverse harmonic}

The transverse modulation was represented by the second harmonic

\begin{equation}
    g_T(\psi)
    =
    \sin 2\psi
    =
    2\sin\psi\cos\psi.
    \label{eq:empirical_transverse_modulation}
\end{equation}

This function has the required twofold periodicity,

\begin{equation}
    \sin\left[2(\psi+\pi)\right]
    =
    \sin 2\psi,
\end{equation}

and changes sign twice during each half-cycle. It also vanishes at the
principal extrema and zeroes of the projected radial factor. It therefore
provides a lowest-order signed transverse harmonic consistent with the
twofold symmetry of the radial model.

The corresponding acceleration is

\begin{equation}
    \mathbf{a}_{T}
    =
    \frac{P_{\odot}}{r^{2}}
    \Gamma_T
    \sin\left[2(f-f_{0})\right]
    \hat{\mathbf{T}}.
    \label{eq:empirical_transverse_acceleration}
\end{equation}

For positive \(\Gamma_T\), this acceleration is directed along the direction
of orbital motion when \(\sin[2(f-f_{0})]>0\), and opposite to the direction
of motion when this quantity is negative.

\subsubsection{Radial--normal--transverse model}

The complete empirical model is

\begin{equation}
\boxed{
\begin{aligned}
    \mathbf{a}_{\mathrm{NG}}^{(RNT)}
    =
    \frac{P_{\odot}}{r^{2}}
    \bigg[
        &\Gamma_R
        |\cos(f-f_{0})|
        \hat{\mathbf{R}}
        \\[2pt]
        &+
        \Gamma_N
        \cos(f-f_{0})
        \hat{\mathbf{N}}
        \\[2pt]
        &+
        \Gamma_T
        \sin\left(2[f-f_{0}]\right)
        \hat{\mathbf{T}}
    \bigg].
\end{aligned}
}
\label{eq:empirical_RNT_model}
\end{equation}

The three empirical models are nested according to

\begin{equation}
    RNT
    \xrightarrow{\Gamma_T=0}
    RN
    \xrightarrow{\Gamma_N=0}
    R.
    \label{eq:empirical_model_hierarchy}
\end{equation}

This nesting permits the improvement associated with each additional
acceleration component to be assessed directly.

\subsubsection{Interpretation and limitations}

The fitted coefficients have dimensions of area per unit mass,

\begin{equation}
    [\Gamma_R]
    =
    [\Gamma_N]
    =
    [\Gamma_T]
    =
    \mathrm{m^{2}\,kg^{-1}}.
\end{equation}

The coefficient \(\Gamma_R\) controls the twofold, outward radial response;
\(\Gamma_N\) controls the signed first-harmonic orbital-normal response; and
\(\Gamma_T\) controls the signed second-harmonic transverse response. Only
\(\Gamma_R\) has an immediate absorption-like interpretation. The normal and
transverse coefficients characterize the acceleration pattern required by
the astrometry but do not uniquely identify its physical origin.

The models are expressed in the instantaneous osculating orbital frame
\((\hat{\mathbf{R}},\hat{\mathbf{T}},\hat{\mathbf{N}})\), rather than in a
spacecraft-fixed body frame. They therefore do not independently determine a
spacecraft pole, surface reflectivity, thermal state, or detailed geometry.
Their purpose is to establish which low-order, phase-dependent acceleration
components are supported by the astrometric observations. More physically
specific cylinder, vector-panel and oblate-spheroid models were subsequently
used to investigate possible geometrical interpretations.
\subsection{Vector Solar-Panel Radiation-Pressure Model}
\label{app:vector_panel_model}

This model represents the nongravitational acceleration produced by a
two-sided planar solar-panel assembly with a fixed orientation in inertial
space. Unlike the empirical radial--transverse--normal models, the
acceleration direction is calculated directly from the instantaneous
Sun--object geometry and an adopted panel-normal vector.

\subsubsection{Geometrical definitions}

Let the heliocentric position of the object be

\begin{equation}
    \mathbf{r}=r\hat{\mathbf r},
\end{equation}

where $\hat{\mathbf r}$ points from the Sun towards the object. The
anti-solar unit vector used in the force model is therefore

\begin{equation}
    \hat{\mathbf s}=\hat{\mathbf r}.
\end{equation}

The fixed panel-normal vector is specified by ecliptic longitude
$\lambda$ and latitude $\beta$:

\begin{equation}
    \hat{\mathbf n}
    =
    \begin{pmatrix}
        \cos\lambda\cos\beta \\
        \sin\lambda\cos\beta \\
        \sin\beta
    \end{pmatrix}.
\end{equation}

In the adopted spacecraft interpretation, $\hat{\mathbf n}$ is identified
with the principal spin-axis direction and with the normal to the plane
of the solar-panel assembly.

The signed incidence cosine is

\begin{equation}
    \mu_0=\hat{\mathbf n}\cdot\hat{\mathbf s}.
\end{equation}

Because the panel is treated as illuminated from either side, an effective
normal pointing away from the illuminated face is defined by

\begin{equation}
    \hat{\mathbf n}_{\mathrm{eff}}
    =
    \begin{cases}
        \hat{\mathbf n}, & \mu_0\geq 0, \\[2mm]
       -\hat{\mathbf n}, & \mu_0<0,
    \end{cases}
\end{equation}

and the non-negative projected-area factor is

\begin{equation}
    \mu=|\mu_0|.
\end{equation}

Thus, $\mu=1$ corresponds to illumination normal to the panel, while
$\mu=0$ corresponds to edge-on illumination.

\subsubsection{Acceleration law}

The implemented acceleration is

\begin{equation}
\boxed{
    \mathbf a_{\mathrm{panel}}
    =
    P_{\odot}
    \left(\frac{1\,\mathrm{au}}{r}\right)^2
    \mu
    \left[
        \Gamma_{\mathrm{panel}}\hat{\mathbf s}
        +
        \Gamma_{\mathrm{panel,n}}\,
        \mu\,\hat{\mathbf n}_{\mathrm{eff}}
    \right]
}
\label{eq:vector_panel_acceleration}
\end{equation}

where $P_{\odot}$ is the solar-radiation pressure at $1\,\mathrm{au}$,
expressed in the units required by the orbit integrator. The two fitted
coefficients have units of area per unit mass:

\begin{equation}
    [\Gamma_{\mathrm{panel}}]
    =
    [\Gamma_{\mathrm{panel,n}}]
    =
    \mathrm{m^2\,kg^{-1}}.
\end{equation}

The first component of Equation~\eqref{eq:vector_panel_acceleration} is

\begin{equation}
    \mathbf a_{\mathrm{dir}}
    =
    P_{\odot}
    \left(\frac{1\,\mathrm{au}}{r}\right)^2
    \Gamma_{\mathrm{panel}}\mu\hat{\mathbf s}.
\end{equation}

It represents the effective anti-solar momentum transfer associated
principally with absorption and direct radiation pressure. Its magnitude
contains one factor of $\mu$ because the illuminated projected area
decreases as the panel approaches an edge-on orientation.

The second component is

\begin{equation}
    \mathbf a_{\mathrm{normal}}
    =
    P_{\odot}
    \left(\frac{1\,\mathrm{au}}{r}\right)^2
    \Gamma_{\mathrm{panel,n}}\mu^2
    \hat{\mathbf n}_{\mathrm{eff}}.
\end{equation}

This term represents an effective reflection-like recoil along the
illuminated panel normal. It contains two factors of $\mu$: one from the
projected illuminated area and one from the angular dependence of the
normal momentum transfer.

The parameters $\Gamma_{\mathrm{panel}}$ and
$\Gamma_{\mathrm{panel,n}}$ are effective dynamical coefficients. They
combine panel area, spacecraft mass and optical response, and therefore
should not be interpreted as independent physical reflectivities without
additional constraints on the panel geometry and spacecraft mass.

\subsubsection{Orbital components}

At each integration epoch, the acceleration is projected onto the
instantaneous orbital radial, transverse and normal triad:

\begin{align}
    a_R &= \mathbf a_{\mathrm{panel}}\cdot\hat{\mathbf R},\\
    a_T &= \mathbf a_{\mathrm{panel}}\cdot\hat{\mathbf T},\\
    a_N &= \mathbf a_{\mathrm{panel}}\cdot\hat{\mathbf N}.
\end{align}

Consequently, transverse and orbital-normal accelerations arise
geometrically even though no independent transverse coefficient is
introduced. In particular, the component parallel to
$\hat{\mathbf n}_{\mathrm{eff}}$ generally projects into all three orbital
directions as the object moves around the Sun.

\subsubsection{Parameter estimation}

For each adopted pole $(\lambda,\beta)$, the six-component orbital state
and the two coefficients

\begin{equation}
    \Gamma_{\mathrm{panel}},
    \qquad
    \Gamma_{\mathrm{panel,n}}
\end{equation}

were adjusted by least squares using the astrometric observations. The
pole coordinates were fixed during each individual fit and were either
adopted from the photometric analysis or selected through raster
scanning.

Formal parameter dispersions were obtained from the covariance matrix of
the converged least-squares solution. They are therefore conditional on
the adopted pole:

\begin{equation}
    \sigma_{\Gamma_i}=\sqrt{C_{ii}},
\end{equation}

with the correlation between the two fitted coefficients given by

\begin{equation}
    \rho_{\mathrm{p,pn}}
    =
    \frac{
        C_{\mathrm{p,pn}}
    }{
        \sigma_{\Gamma_{\mathrm{panel}}}
        \sigma_{\Gamma_{\mathrm{panel,n}}}
    }.
\end{equation}

These formal uncertainties do not include uncertainty in the
raster-selected pole or systematic uncertainty associated with the
assumed force law.

\subsubsection{Model assumptions and limitations}

The implemented vector-panel model assumes:

\begin{enumerate}
    \item a fixed principal-axis pole in inertial space;
    \item a panel normal parallel to that pole;
    \item identical effective optical behaviour on both panel faces;
    \item a single equivalent planar area representing both solar wings;
    \item inverse-square scaling with heliocentric distance;
    \item no bus contribution to the radiation force;
    \item no mutual shadowing or occultation between the bus and panels;
    \item no thermal lag or Yarkovsky recoil;
    \item no nutation or time-dependent attitude motion; and
    \item no separate modelling of absorption, specular reflection and
          diffuse reflection.
\end{enumerate}

Rotation about the panel normal does not alter
Equation~\eqref{eq:vector_panel_acceleration}; consequently, explicit
spin-phase averaging is unnecessary under these assumptions. The model
should therefore be regarded as an intermediate physical approximation
between the empirical orbital-component models and a complete
spacecraft macromodel.

\end{document}